\documentclass[linenumbers]{aastex631}

\accepted{November 10, 2022}

\begin{document}

\title{Successive interacting  coronal mass ejections: How to create a perfect storm?}

\author[0000-0003-3397-7769]{{G. J. Koehn}}
\affiliation{Blackett Laboratory, Imperial College London, London, UK}

\author[0000-0002-2015-4053]{R. T. Desai}
\affiliation{Centre for Fusion, Space and Astrophysics, University of Warwick, UK}
\affiliation{Blackett Laboratory, Imperial College London, London, UK}

\author[0000-0001-9992-8471]{E. E. Davies}
\affiliation{University of New Hampshire, New Hampshire, USA}

\author[0000-0003-2701-0375]{R. J. Forsyth}
\affiliation{Blackett Laboratory, Imperial College London, London, UK}

\author[0000-0003-4733-8319]{J. P. Eastwood}
\affiliation{Blackett Laboratory, Imperial College London, London, UK}

\author[0000-0002-1743-0651]{S. Poedts}
\affiliation{Katholieke Universiteit Leuven, Leuven, Belgium}
\affiliation{Institute of Physics, University of Maria Curie-Sk{\l}odowska, Lublin, Poland}



\begin{abstract}

{Coronal mass ejections (CMEs) are the largest type of eruptions on the Sun and the main driver of severe space weather at the Earth. In this study, we implement a force-free spheromak CME description within 3-D magnetohydrodynamic simulations to parametrically evaluate successive interacting CMEs within a representative heliosphere. We explore CME-CME interactions for a range of orientations, launch time variations and CME handedness and quantify their geo-effectiveness via the primary solar wind variables and empirical measures of the disturbance storm time index and subsolar magnetopause standoff distance. We show how the interaction of two moderate CMEs between the Sun and the Earth can translate into extreme conditions at the Earth and how CME-CME interactions at different radial distances can maximise different solar wind variables that induce different geophysical impacts. In particular, we demonstrate how the orientation and handedness of a given CME can have a significant impact on the conservation and loss of magnetic flux, and consequently B$_z$, due to magnetic reconnection with the interplanetary magnetic field. This study thus implicates identification of CME chirality in the solar corona as an early diagnostic for forecasting geomagnetic storms involving multiple CMEs.   }

\end{abstract}



\section{Introduction}

Coronal mass ejections (CMEs) are the largest type of eruption seen from the solar atmosphere and are the primary cause of severe geomagnetic storms and disturbances when they arrive at the Earth. While most geomagnetic storms are created by the impact of a single CME, over a quarter are caused by the interaction of multiple CMEs \citep{Zhang2007}; and it has been suggested that the majority of fast complex ejecta in the solar wind are produced by multiple CMEs \citep{Burlaga2001}. Several studies have examined multiple successive interplanetary CMEs (ICMEs) and determined that preconditioning of the solar wind, where an initial CME clears the path for a second CME \citep{Liu2014,Temmer15,Desai2020}, and CME--CME collisions \citep{Shiota2016,Scolini2020}, can result in a significant increase in geo-effectivness, i.e. the severity of the impact on Earth. Quantification of this amplification has the potential to address the question of what upstream conditions can produce an exceptionally geo-effective event, i.e. a `perfect storm', at the Earth. 

\hspace{2mm} A survey of extreme space weather events reveals that the most severe events on record have featured multiple successive CMEs. For example, the famous Carrington event of 1859 involved multiple successive eruptions and four nights of low-latitude aurora \citep{Tsurutani2003}, and the earlier solar storm of September 1770 was reported to consist of repeated low-latitude aurora for nearly 9 days \citep{Hayakawa17}. Early events in the Space Age also exhibited this phenomenon, including the August 1972 solar storm which resulted in the record Sun-1~AU CME transit time of 14.6 hours \citep{Knipp18}, and the solar storm of March 1989, which triggered widespread power grid failures in North America and Europe \citep{Boteler19}. Further notable storms in 2003 \citep{Webb04} and 2017 \citep{Scolini2020} similarly featured multiple CMEs. 

For the past decade, in space weather operations, CMEs have mainly been modeled using the cone model, which has been shown to accurately capture their shock arrival times \citep{Odstrcil1999,Xie2004,Xue2005,Chane06}. This approach, however, neglects the intrinsic magnetic field structure and thus fails to predict the magnetic characteristics at the Earth or capture \{the magnetised interactions of a CME with another. More enhanced CME descriptions include an intrinsic toroidal flux rope structure, adding complexity in defining initial conditions \citep{Gibson1998,Titov1999,Jin2017,Torok2018}. Recent studies have taken simpler approaches to simulate flux ropes in the heliosphere utilising a spheromak approximation of a CME \citep{Vandas97,Vandas98} initialised outside of the solar corona where they can be prescribed initial conditions commensurate with cone-based models \citep{Kataoka2009,Shiota2016,Verbeke2019,Singh2020a}. In this study we thus utilise the spheromak CME description to theoretically evaluate the direct interaction of successive ICMEs and assess how this can enhance CME geo-effectiveness.

This article is structured as follows: Section~\ref{MHD} provides a description of the simulation approach employed to model the solar wind and CMEs, as well as detailing the methods employed to constrain geo-effectiveness. Section~\ref{sec:ParametricStudy} contains the design and analysis of the parametric study simulations, followed by a discussion of the results and future prospects in Section~\ref{sec:Conclusion}.
  
\section{Magnetohydrodynamic Modelling}
\label{MHD}
\subsection{Heliospheric Set-up}
In this study we use a magnetohydrodynamic (MHD) heliospheric model developed from the astrophysical PLUTO code. A description of the heliospheric model is provided by \citet{Desai2020} and the underlying numerics are presented by \citet{Mignone2007}. 
In this study, we employ the Harten-Lax-van-Leer discontinuities (HLLC) approximate Riemann solver which is stable for highly magnetised plasmas \citep{David2019}. 
In this study we self-consistently model the solar wind evolving through the inner heliosphere as well as the insertion and evolution of multiple CMEs. 

The ideal MHD equations are solved across a physical domain extending from 0.1~AU ($\approx$21.50 solar radii ($R_S$)) out to 1.1~AU ($\approx$~236.5~$R_s$), colatitudes $\theta$ = 30$^{\circ}$ to 150$^{\circ}$ and longitudes $\phi$ = -60$^{\circ}$ to 60$^{\circ}$. This reduction of the global heliosphere to a solid angle volume allows us to capture all dynamics of the CMEs focused on herein, whilst reducing the computational requirements. The static computational grid has 60 and 273 cells along angular and radial axes, respectively, which produces a radial spatial resolution of 0.788 $R_S$ and angular resolution of $2^{\circ}$. This set-up was validated with an error convergence analysis conducted up to a resolution 3.3x greater than this (not shown). The $\phi$-boundaries, defined by the coordinate range above, implement periodic boundary conditions to represent a closed system in azimuth.

 \subsection{Solar Wind \& Parker Spiral}
 To produce a heliosphere we implement a slow solar wind at the inner boundary of 0.1~AU to represent outflow from the streamer belt located around the Sun's equator, and an interplanetary magnetic field (IMF) structure based upon the \citet{Parker1958} model. The solar wind mostly comprises electrons and protons, but also some heavier elements, which are reflected in a mean molecular mass of $\mu =0.6$. The model assumes the radial speed to be only a function of radius, Faraday's law and Maxwell's equation $\nabla \mathbf{\cdot B} = 0$ for a spherically symmetric geometry, leading to the explicit magnetic field components:
\begin{equation}\label{eqn:radialInnerBoundaryConditions_Br}
    B_r =  B_{S}  \Big(\frac{R_{S}}{r} \Big)^2,~~~~~~~~
    B_{\theta} = 0,~~~~~~~~
    B_{\phi} = -B_{S} \frac{\omega_{S} R_{S}}{v} \frac{R_{S}}{r} \sin \theta,
\end{equation}
where the magnetic field at the solar surface, $B_S$ = 2.5$\cdot$10$^5$~nT and solar rotation rate, $\omega$~=~2.7$\cdot$10$^{-6}$~rad~s$^{-1}$. At the inner boundary (0.1\,AU) the radial velocity and density are set to 300\,km/s and 666.66\,m$_p$/cm$^3$, respectively. 
 
\begin{figure}[ht]
    \begin{center}
            \includegraphics[width=0.6\textwidth]{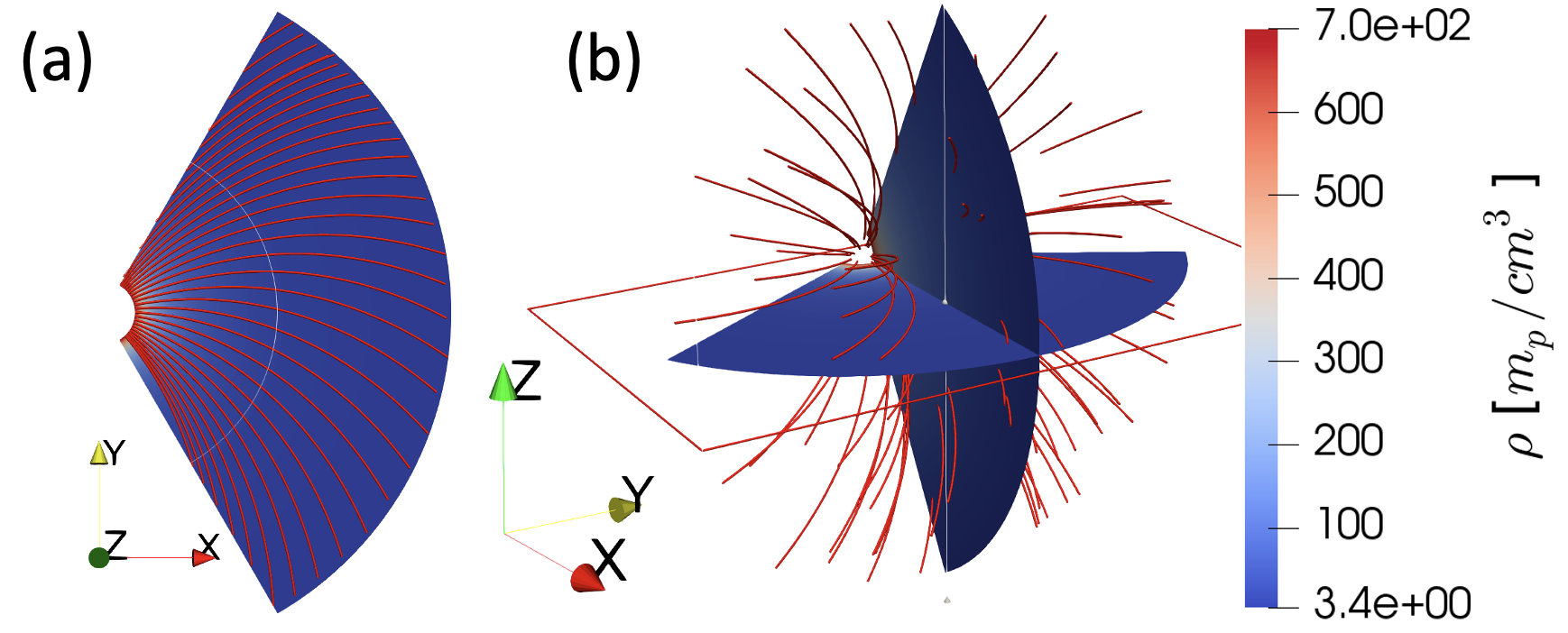}
    \end{center}
    \caption{Density profile and magnetic field lines (red) in the steady-state heliosphere from (a) the equatorial plane and (b) showing both the meridional and equatorial planes. }
    \label{fig:2and3DParkerSpiral}
\end{figure}

 In order to establish a self-consistent solution to the solar wind within the simulation domain, a simulation is set to run for 10 days. After $\approx$7 days a steady state solution is established with values at 1~AU of $v_r$ = 477.66~km/s, density, $\rho$ = 4.21~m$_p$/cm$^3$, B$_r$ = 5.46~nT, B$_{\theta}$ = 0.0001~nT and $B_{\phi}$ = -2.57~nT. These values are well within acceptable conditions for a slow solar wind and the resulting magnetic field structure shown in Figure \ref{fig:2and3DParkerSpiral} clearly shows an Archimedean \citet{Parker1958} spiral wrapping clockwise around the Sun when viewed from above.

\newpage
\subsection{Spheromak CME approximation}\label{subsec:Spheromak CME Implementaiton}
To model CMEs, we utilise a linear force-free spheromak (LFFS) \citep{Chandrasekhar1957,Verbeke2019} to capture the intrinsic magnetic field flux rope-like structure. Spheromak CMEs can be thought of as an approximation of the main frontal lobe of a CME but assume spherical closure which is in contrast to observations where CMEs are often observed to possess ellipsoidal or croissant-like shapes with footpoints still connected to the Sun \citep{Janvier_2013}. The model is therefore more appropriate for frontal impacts as opposed to glancing blows~\citep{Maharana2022}.

The implementation and conservation of this complex magnetic field topology through the inner boundary is non-trivial due to magnetic field divergence considerations and in this implementation the spheromak is initialised over multiple grid cells near the inner boundary to ensure numerical stability. The spheromak defines a spherical region of space and the hydrodynamic quantities are thus commensurate with those implemented for a cone based model with the exception of the velocity which features an additional expansion due to the magnetic pressures. The magnetic fields of the spheromak CME are expressed as 
\begin{equation}\label{eqn:radialInnerBoundaryConditions_Br0}
    B_{r^\prime} =  2 B_{0}  \Big(\frac{j_{1}(\alpha r^\prime)}{\alpha r^\prime} \Big) \cos\theta^\prime,~~~~~~~~   
    B_{\theta^\prime} =  - B_{0}  \Big(\frac{j_{1}(\alpha r^\prime)}{\alpha r^\prime} + j_{1}(\alpha r^\prime) \Big) \sin\theta^\prime,~~~~~~~~
    B_{\phi^\prime} = H \cdot B_0 j_1(\alpha r^\prime) \sin\theta^\prime.
\end{equation}
Here the coordinate system, r$^\prime$, $\theta^\prime$ and $\phi^\prime$, is defined by the main axis of the spheromak and j$_1$($\alpha r^\prime$) is the vanishing spherical Bessel function of order unity with $\alpha$r$_0 \approx 4.48$ to satisfy the force-free condition with the magnetic field tending to zero at the boundary of the spheromak. The handedness, H, defines the direction of toroidal flux and can take values of $\pm$1. 

To demonstrate and verify the implementation within the heliospheric model, the spheromak was initialised within a local simulation with plasma conditions representative of 0.1~AU. These conditions were chosen to closely resemble the conditions close to the inner simulation boundary, at 0.1~AU, of the subsequent parametric study. The input parameters correspond to those used in the subsequent parametric study as listed as CME1 in Table \ref{tab:ParameterStudy}, and a grid resolution of 0.75~$R_S$ radial, 9$^\circ$ azimuthal, and 4$^\circ$ polar was used with a second order Runge-Kutta algorithm for tracing the magnetic field lines. Figure~\ref{fig:spheromakVerification} shows the spheromak shortly after initialisation and then after 60 minutes have elapsed. The structure of the spheromak is well-preserved although differences can be seen which are attributed to numerical diffusion and magnetic expansion. The conservation of the global structure demonstrates the force-free implementation of the spheromak.
  
\subsection{Quantifying Geo-effectivness}\label{subsec:Quantifying geo-effectivness}

To diagnose the results of Sun-to-Earth CME simulations we insert a virtual spacecraft into the simulation domain in Figure~\ref{fig:2and3DParkerSpiral} at X = 1~AU, at colatitude $\theta=90^\circ$ and longitude $\phi= 0^\circ$. Multiple latitudes and longitudes close to the Sun-Earth line were however examined to ensure the results were representative (not shown). This virtual spacecraft outputs timeseries of all modeled variables and allows us to identify interplanetary CME signatures and indicators of geo-effectiveness. The primary controllers of geo-effectiveness stem from the solar wind velocity and density, which controls the pressure exerted on the magnetosphere and the southward component of the IMF, B$_Z$, which drives magnetic reconnection and the \citet{Dungey1961} cycle.

\subsubsection{Disturbance Storm Time Index}

To provide a more sophisticated estimate of geo-effectiveness based upon these key solar wind variables, we use the empirical relationship determined by \citet{Wu2005} for the minimum Disturbance Storm Time (Dst) index induced by a given CME. This provides a geo-effectiveness estimate based on the ring current content in response to the solar wind B$_Z$ and v$_r$. This empirical relationship is based on measurements of 135 magnetic clouds, defined as regions of enhanced magnetic field, a smooth rotation of the magnetic field's direction and a low plasma beta \citep{Burlaga1981,Lepping1990}. This Dst relationship is written as:
\begin{equation}\label{eqn:DstWu}
   Dst_{min} = -16.48-12.89 \cdot (v_r \cdot B_Z)_{max}, 
\end{equation}
where $B_Z$ =  $|B_{GSM_z}|$ for $(B_{GSM})_z$ $<$ 0 and $B_Z$ = 0 for $(B_{GSM})_z$  $\geq$ 0. $(B_{GSM})_z$ is thus equal to $(B_{GSE})_z$ as outputted from the simulations for the case of zero dipole tilt. Note that this cap of Dst is just required for this relationship: the Dst can be positive, yet cannot be estimated with this relationship in the positive regime. It shall be mentioned here that this relationship has been found to have an error of less than 10\% for magnetic cloud regions, yet larger errors for CME sheath regions. 
This Dst relationship can be used to categorise storms as either weak ($\leq$ --30 nT), moderate ($\leq$ --50 nT), strong ($\leq$ --100 nT), severe ($\leq$ --350 nT), and great or extreme ($\leq$ --350 nT), according to the commonly-used classifications \citep{Loewe1997}. 
While this model estimates the minimum Dst, this variable is plotted over time to understand the location of the minimum Dst. It is also based upon the solar wind electric field which itself represents a solar wind-magnetosphere coupling function \citep{Newell15}. In addition to the extrema, the timeseries thus also illustrates the time-dependent nature of the solar wind-magnetosphere coupling.

\subsubsection{Magnetospheric Compression}
The complexity of the solar wind-magnetosphere interactions leads to a myriad space weather impacts, which are complex to quantify with a single parameterisation. We thus also utilise the \citet{Shue1998} model to predict the location of the subsolar terrestrial magnetopause. The magnetopause controls the outer boundary of the magnetosphere and its motion is a fundamental driver of outer radiation belt dynamics. Moreover, rapid compression inside of geosynchronous orbits can accelerate new radiation belts across sub-drift time-scales due to the generation of a compressive front which propagates through the magnetosphere at speeds compariable to particle drift speeds. This process led to the highest observed electron energies of $>$20~MeV in the inner magnetosphere following the interplanetary shock of March 1991. \citep{Blake1992,Horne15}. This is therefore used as a further diagnostic of the magnetospheric and radiation belt response. The \citet{Shue1998} model is based upon the solar wind dynamic pressure, D$_p$ and B$_Z$, and the subsolar magnetopause is expressed as
\begin{equation}
    r_{MP_{SS}} =  (10.22 + 1.89 \textrm{ tanh}[0.184 (B_Z + 8.14)])(D_p)^{-1/6.6}. 
\end{equation}
It should be noted that large deviations have been reported inside 8 R$_E$ \citep{Staples2020}, but the underlying dependence on dynamic pressure provides a consistent estimate across the events studied herein. Interplanetary shocks at the leading edge of CMEs can also trigger large-scale oscillations of the magnetopause surface which result in large overshoots \citep{Freeman1995,Desai2021,Cahill92} which might also cause the magnetopause to be temporarily compressed closer to the Earth than indicated by this steady-state prediction.

\subsection{Parametric Study}\label{sec:ParametricStudy}
To efficiently progress through a large possible space of CME-CME interactions, the parametric study was conducted linearly in three stages, each time choosing the most geo-effective case and moving to the next stage. The baseline scenario has been chosen to be two CMEs originating from the same active region and propagating along the centre of the domain outward. The tilt angle of a single spheromak around the Heliocentric Earth EQuatorial (HEEQ) $x$-axis was first chosen to be varied within \emph{Stage~1}. The tilt angle changes the axis of symmetry of the spheromak (0$^\circ$ aligned with HEEQ $z$-axis, 90$^\circ$ aligned with HEEQ $y$-axis). Once the most geo-effective orientation was identified, \emph{Stage 2} proceeded to model two successive CMEs to examine the effects of CME-CME collisions. The next parameter varied was the waiting time between the two launches to produce varying CME interaction dynamics. Finally, the handedness of the CMEs were varied in \emph{Stage 3}, in order to further explore magnetic interactions between two CMEs.

\subsection{Input Parameters of Spheromak}\label{subsec:InputParameters}
The initial conditions of CME parameters can be constrained from remote-sensing observations of CMEs and their source region on the Sun. 
The choice of parameters here is based on prior estimation made by \citet{Scolini2019} for the CME-CME event recorded on the 13/14 June 2012. The choice was thus made to inject two identical CMEs of different velocities one after another. The initial CME, hereafter referred to as \emph{CME1}, was given a moderate radial speed $v_{radial}$ = $v_r$ = 500~km/s and the subsequent CME, hereafter referred to as \emph{CME2}, would be given a high radial speed of $v_{r}$ = 1500~km/s. When considering magnetic expansion, these speeds correspond to propagation speeds at the front of the CME of 723~km/s and 1,723~km/s, respectively. The precise parameters chosen are outlined in Table~\ref{tab:ParameterStudy}.

\section{Results}
 
\subsection{Single CME}\label{subsec:VerificationModel}

CME1 is first individually simulated with a tilt angle $\tau = 180 ^\circ$ and handedness $H$ =+1. The global simulation outputs of density, radial velocity and southward IMF component, B$_Z$ are shown in Figure~\ref{fig:SingleCMEVerification_Tau180} and the timeseries for the virtual spacecraft at 1~AU is shown in Figure~\ref{fig:Verification_180Timeseries}.
A fast-forward shock is visible at the leading edge of the CME and a high density sheath region follows which contains enhanced magnetic fields components due to a pile-up of the ambient solar wind. 
The shock is thus separated from the flux rope CME where the magnetic field increases and the density falls of to a very low depression compared to the background solar wind level. The CME exhibits a plasma beta below unity, $\beta<1$, between 62~h and 74~h, where plasma motion is dominated by the magnetic field. The magnetic field lines within the CME cause the temperatures to evolve separately from the ambient solar wind and the temperatures within the CME are therefore of the right magnitude of less than 10$^{6}$~K. Importantly, the magnetic field inside the CME exhibits a smooth rotation indicative of a flux rope and which completes the definition of a magnetic cloud~\citep{Burlaga1981}.

The estimated Dst shows a significant depression in the region of the CME and reaches a minimum of -63~nT. This event may therefore be classified as a moderate geomagnetic storm.
The subsolar magnetopause distance is predicted to be compressed to 8~R$_E$ following the initial impact, followed by an expansion.

\subsection{Single CME Stage 1: Tilt Angle}
\label{Stage1}
\begin{figure}[ht]
    \centering
    \includegraphics[width=0.65\textwidth]{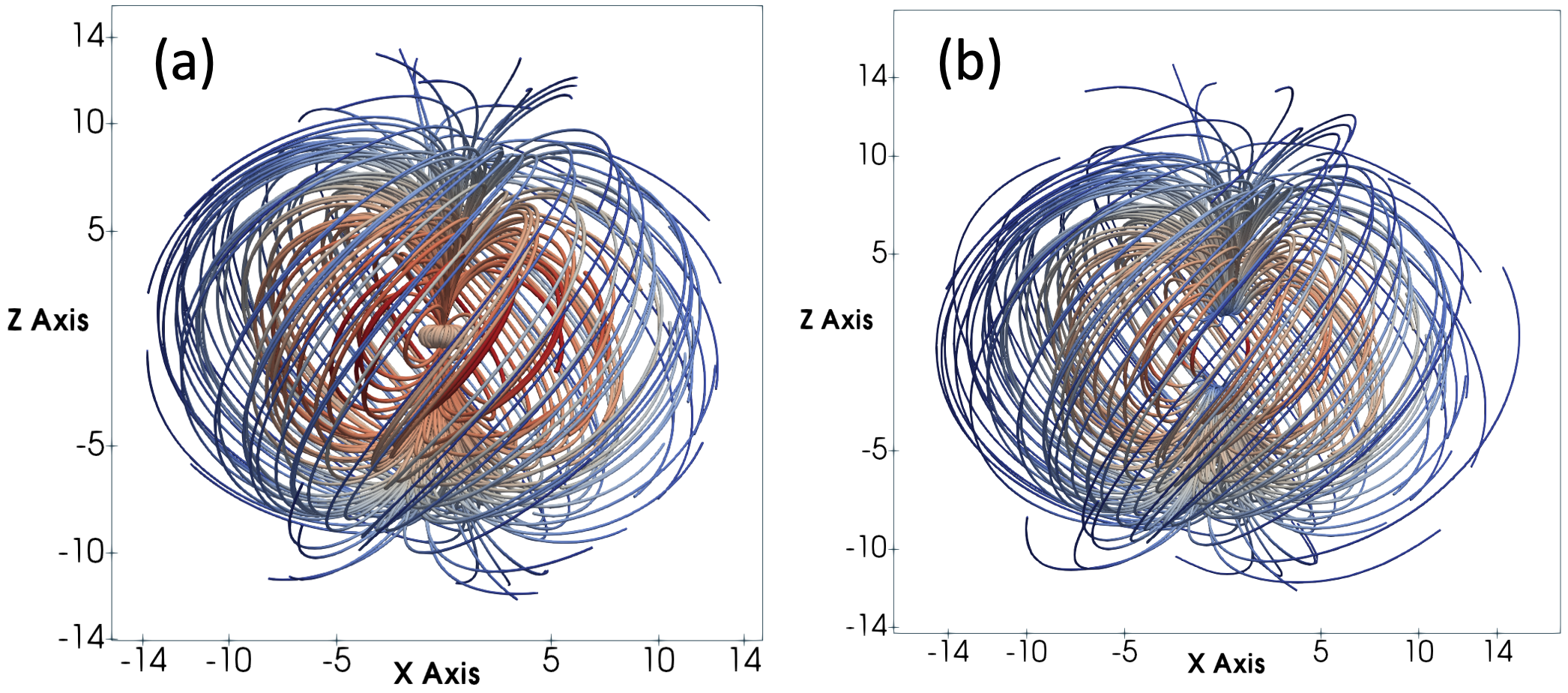}
    \caption{Magnetic field lines of (a) the initialised spheromak and (b) after 60 min of evolution, in static heliospheric conditions representative of conditions 0.1~AU. The colour is proportional to the normalised magnetic field magnitude.}
    \label{fig:spheromakVerification}
\end{figure}

\begin{figure}[ht]
   \centering
  \includegraphics[width=0.7\columnwidth]{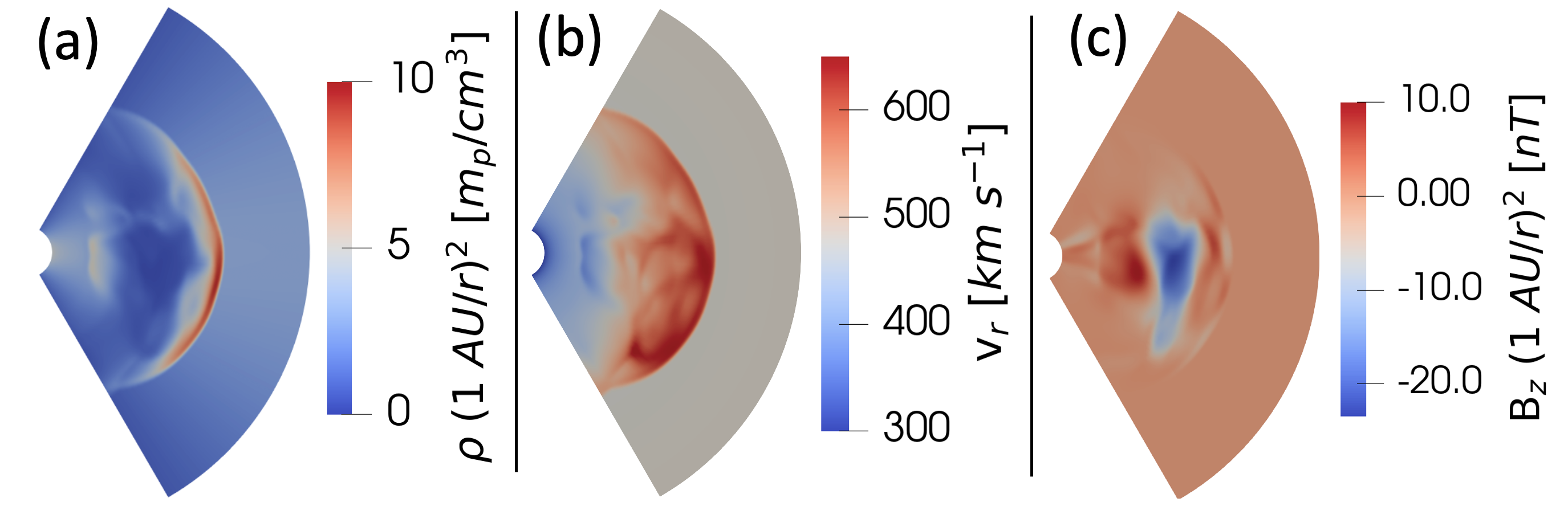}
   \caption{
   CME1 Spheromak simulation with initial tilt angle $\tau$ = 180$^\circ$ and handedness $H$ = +1, 30~h after launch. (a) Shows the normalised density, (b) the radial velocity and (c) (B$_{GSE}$)$_z$ in the equatorial plane. Further CME1 parameters are specified in Table \ref{tab:ParameterStudy}. }\label{fig:SingleCMEVerification_Tau180}
\end{figure}

\begin{figure}[ht]
    \centering
    \includegraphics[width=0.9\columnwidth]{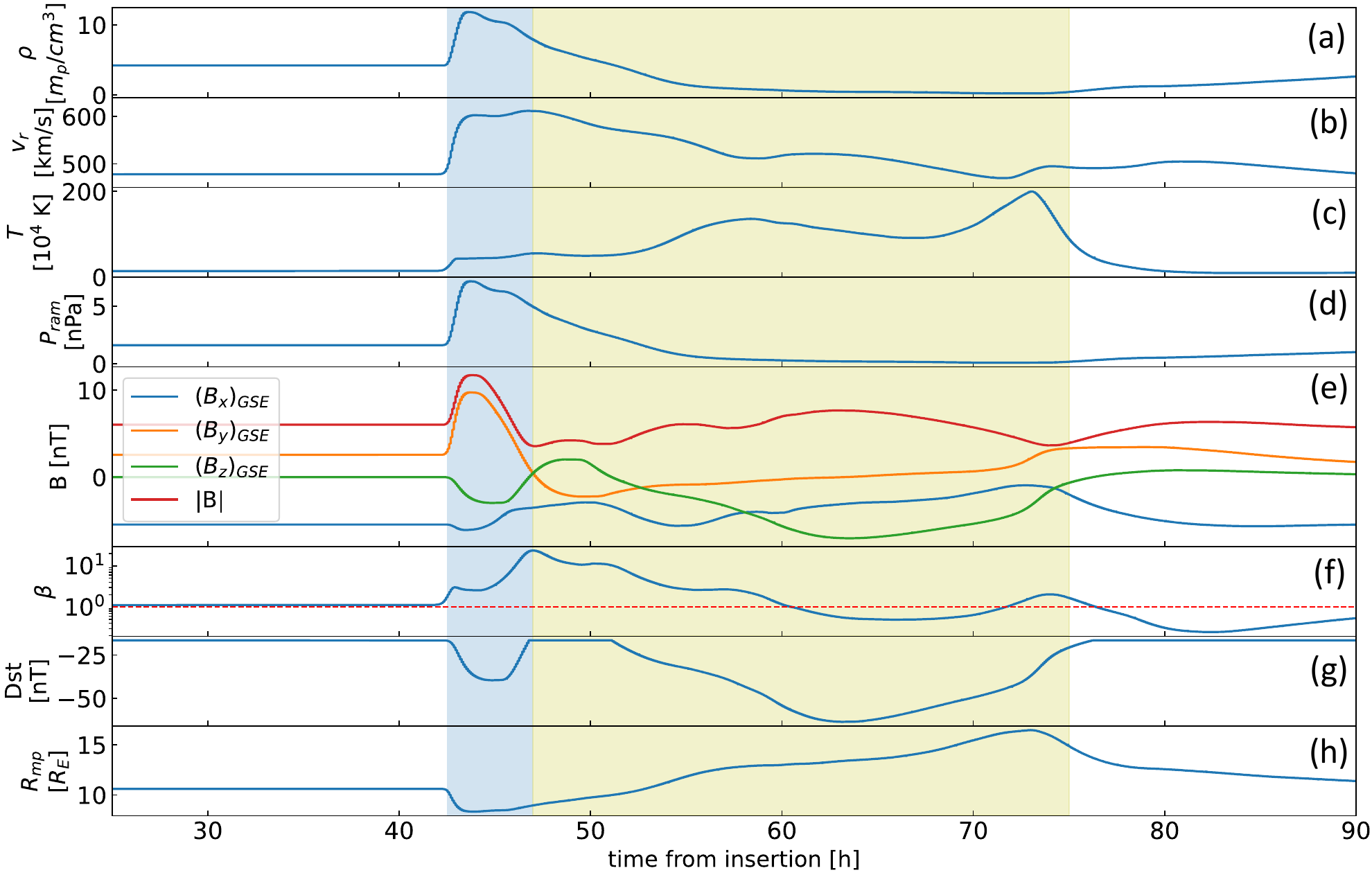}
    \caption{Virtual spacecraft timeseries data at 1~AU of CME1 shown in Figure \ref{fig:SingleCMEVerification_Tau180}. CME1 has a tilt angle $\tau$ = 180$^{\circ}$ and handedness $H$ = +1. Further parameters for CME1 are specified in Table \ref{tab:ParameterStudy}. The blue shaded region indicates the region of the shock and sheath and the yellow shaded region the ejecta/flux rope.}\label{fig:Verification_180Timeseries}
\end{figure}

\begin{figure}[ht]
  \centering
  \includegraphics[width=0.9\columnwidth]{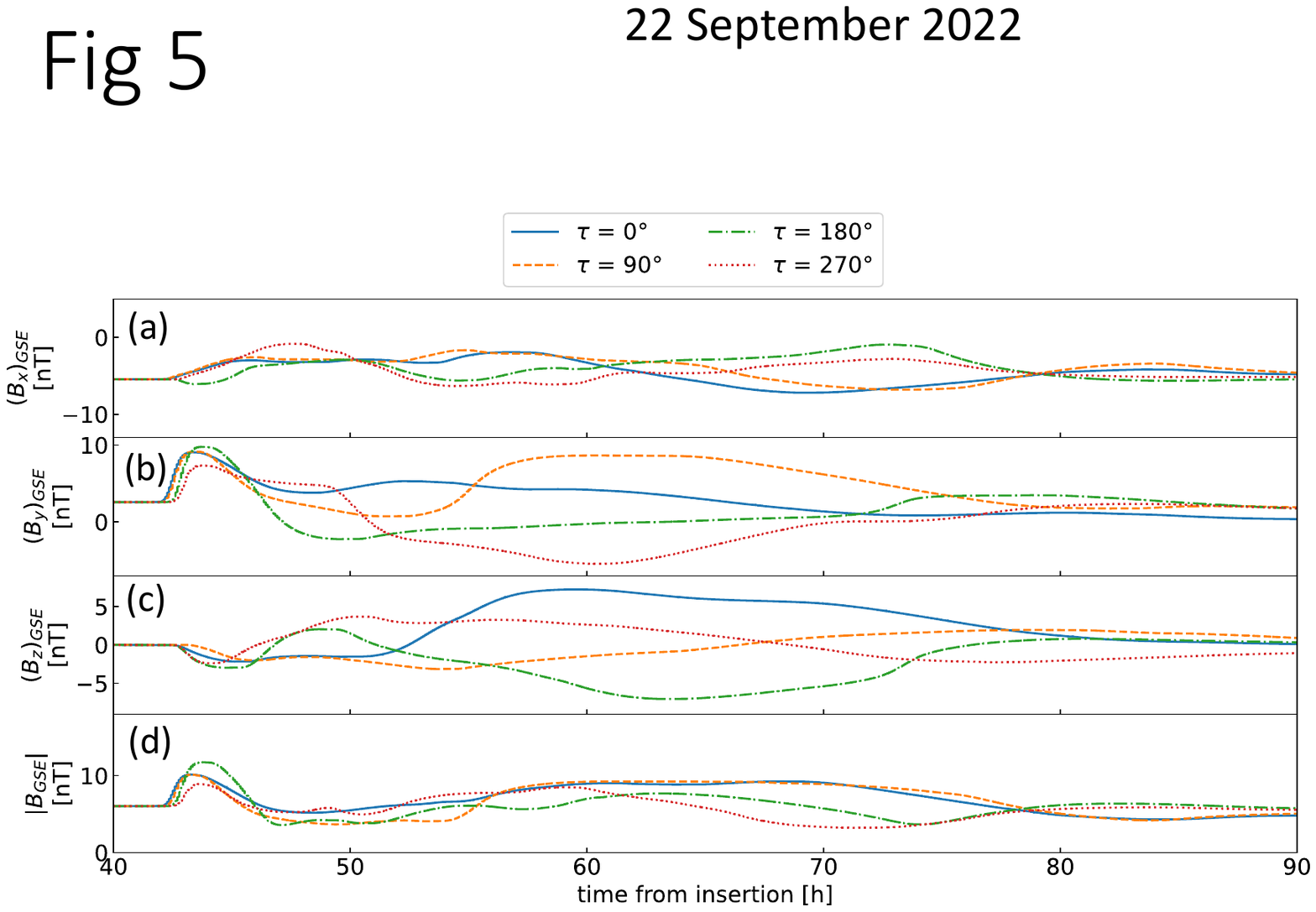}   \caption{\small 
  Magnetic field timeseries at 1 au for CME1 for different initial tilt angles. 
  }\label{fig:Tilt_Bcomponents_Summary}
\end{figure}

The first stage of the parametric study identified the most geo-effective single CME tilt angle. The tilt angle describes the main axis of the spheromak, i.e. +Z in Figure~\ref{fig:spheromakVerification}. To this end, the magnetic field timeseries of four simulations are shown in Figure~\ref{fig:Tilt_Bcomponents_Summary} with tilt angles $\tau$ = 0$^\circ$, 90$^\circ$, 180$^\circ$, 270$^\circ$ and fixed handedness $H$~=~+1. A further two cases of $\tau$ = 135$^\circ$ and 225$^\circ$ are also simulated to further resolve the primary peak/trough in the trends.
The magnetic field structure of the spheromak is strongly deformed in the propagation from Sun to Earth, yet the magnetic field direction of the initial spheromak in the z$'$-axis can still be recovered at 1~AU.

For the $\boldmath{\tau= 0^{\circ}}$ tilt run, the spheromak is inserted in upright orientation as shown in Figure~\ref{fig:spheromakVerification}. Thereby, one would expect the main direction of the magnetic field lines to be in the positive $z$-direction apart from the very centre of the spheromak. A strong positive $(B_{GSE})_z$ component is indeed observable in the virtual spacecraft timeseries between 54~h and 80~h. The $(B_{GSE})_x$ component undergoes a rotation from positive to negative, as expected for the radial component inside the spheromak. The $(B_{GSE})_y$ component however displays some deviation to the spheromaks structure at 0.1~AU.

\begin{figure}[ht]
  \centering
  \includegraphics[width=0.5\columnwidth]{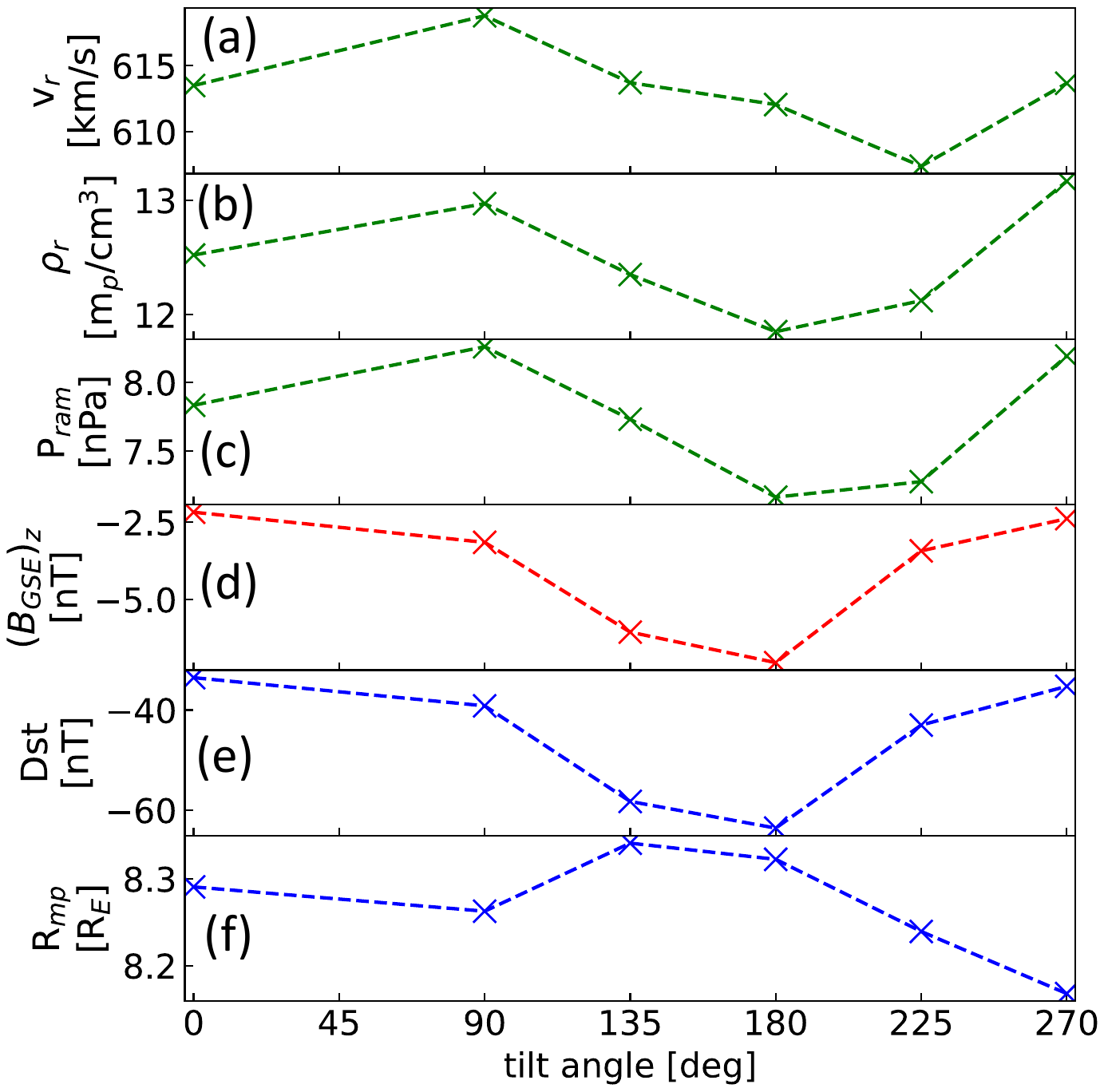}
  \caption{\small 
    Maximal values for the range of tilt angles CME1 is simulated for. (a) shows v$_r$, (b) shows the density, $\rho_r$, (c) the dynamic pressure, P$_{ram}$, (d) $(B_{GSE})_z$, (e) the predicted Dst and (f) the predicted subsolar magnetopause standoff distance, R$_{mp}$.
  } \label{fig:Tilt_Summary}
  \label{fig:dhjgfhd}
\end{figure}

The $\boldmath{\tau= 180^{\circ}}$ tilt run, as presented in Figures~\ref{fig:SingleCMEVerification_Tau180} and \ref{fig:Verification_180Timeseries}, displays the converse trends to this which lead to a large and prolonged southward-directed magnetic field between 50~h and 74~h. A significant twist in the $\phi$-direction is also visible and appears coherent with the negative $B_{\phi}$ component of the Parker spiral.
Analogous trends are visible for for the $\boldmath{\tau= 90^{\circ}}$ and $\boldmath{\tau= 270^{\circ}}$ cases.
The maximum in the selected plasma dynamical variables are shown in Figure \ref{fig:Tilt_Summary}, and indicate the tilt angle of $\tau= 180^{\circ}$ leads to the largest geo-effectiveness.
 
\subsection{CME-CME Stage 2: Waiting Time}\label{sec:WaitingTime}
The second stage of the parametric study proceeded to investigate the interaction of two CMEs at different locations within the simulation domain. The second CME is initialized with a higher velocity of 1,500~km/s compared to the first CME's velocity of 500~km/s and the second CME is thus gaining quickly on the first CME, see Table~\ref{tab:ParameterStudy}. The waiting time was varied from 12-36~h. This range encompasses the two boundary cases of 1) a collision at $\sim$0.5 AU and 2) a pure preconditioning of the solar wind by the first CME where they subsequently collide beyond the Earth.~\citep[e.g.][]{Temmer2015}
Here a collision of the flux ropes and sheaths of the two CMEs is dubbed as a full collision, unless otherwise specified. 
All spheromaks were set to have an initial tilt angle of $\tau =180^{\circ}$, expecting this would lead to the highest geo-effectiveness as found in Stage 1, and a handedness $H$ = +1.
A total of 7 cases (plus CME1 and CME2 individually) have been run leading to a waiting time resolution of 4~h. Individual simulations for CME1 and CME2 allow us to compare the enhancement induced through the interaction. To elucidate key features of CME-CME interactions, in the following sections we describe two distinct cases in detail which correspond to; 1) a full merge of the two CME including their sheaths (20~h waiting time) and a merge of the magnetic centres only (28\,h waiting time), each occurring at approximately 0.9~AU just ahead of the Earth.

\subsubsection{20 hours: Shock-CME Interaction}
\begin{figure}[ht]
    \begin{center}
        \includegraphics[width=0.9\textwidth]{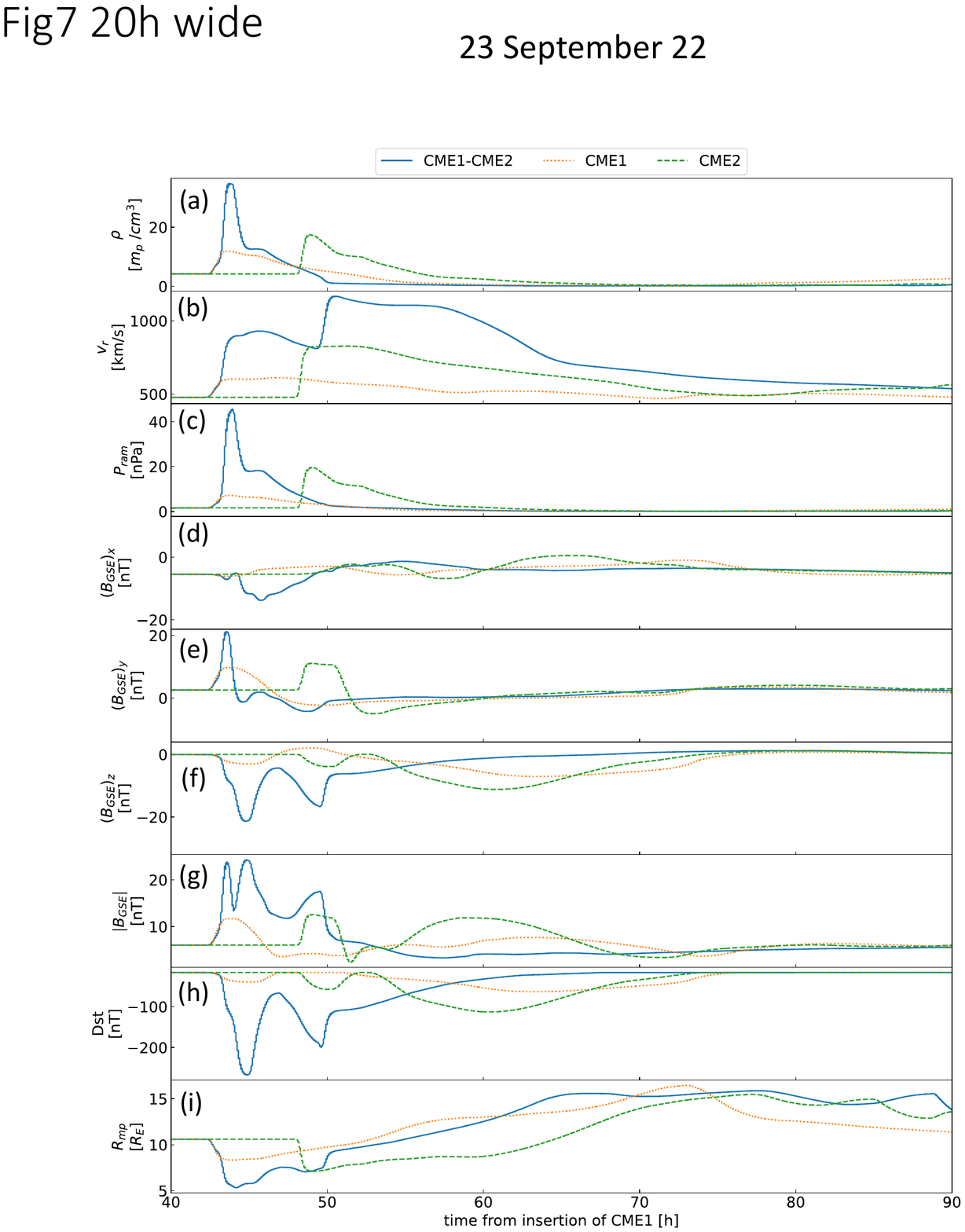} 
    \end{center}
    \caption{Virtual spacecraft data at for the CME1-CME2 interaction with waiting time of 20~h between their respective launches. This corresponded to a merge of the CME sheaths at approximately 0.9~AU. The timeseries in solid blue, dotted yellow and dashed green refer to the signatures of CME1-CME2, CME1, and CME2, respectively.}
    \label{fig:waitingTime20}
\end{figure}

\begin{figure}[ht]
    \begin{center}
            \includegraphics[width=0.8\textwidth]{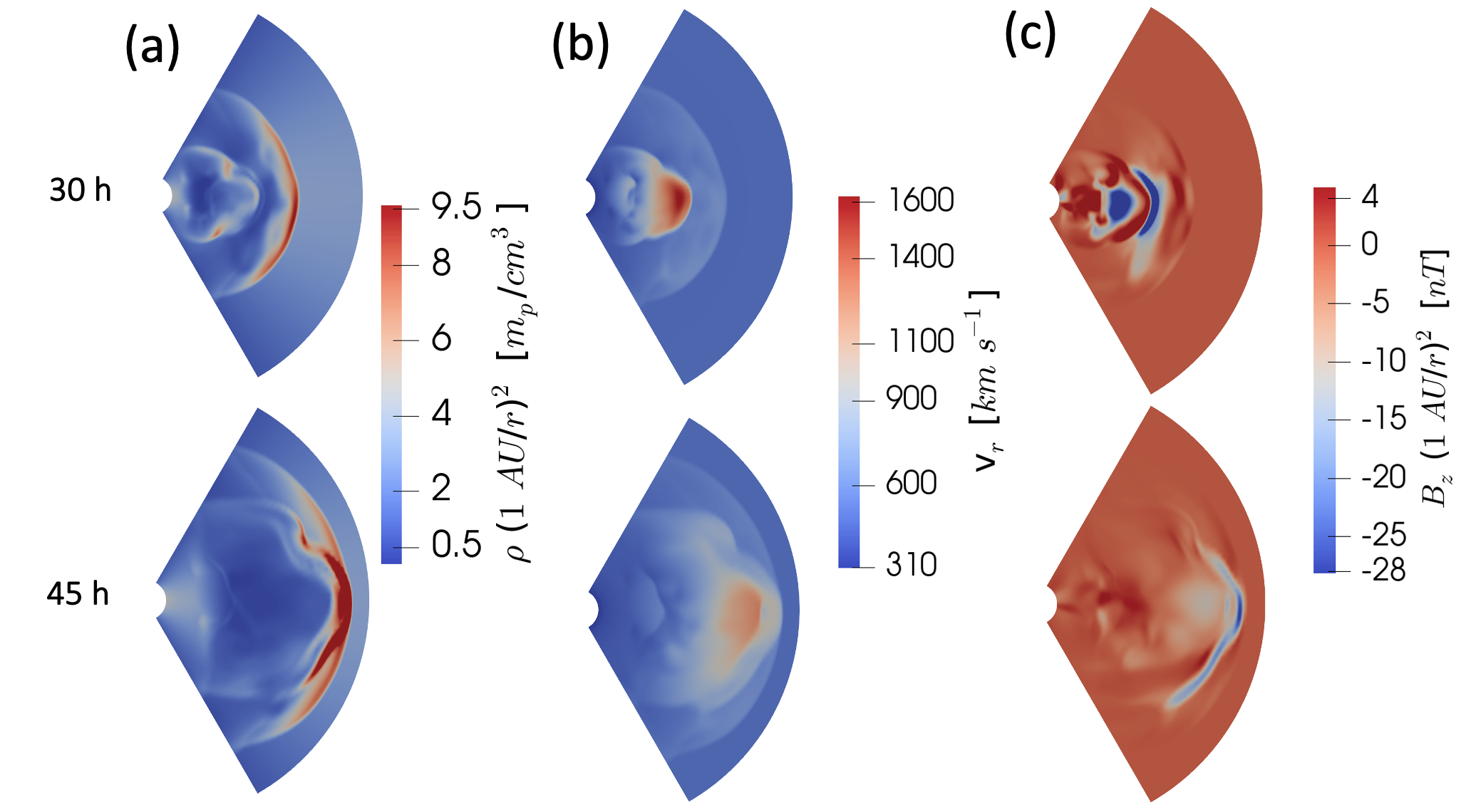}
    \end{center}
    \caption{ CME1-CME2 interaction with waiting time of 20~h between their launches. (a) shows the normalised density, (b) the radial velocity and (c) the (B$_{GSE}$)$_z$ in the equatorial plane.}
    \label{fig:waitingTime20hParaview}
\end{figure}

The waiting time of 20\,h resulted in an extended interaction of both the CME magnetic flux ropes and sheath regions prior to reaching Earth's position. The sheath regions of the two CMEs merge at $\sim$0.9~AU. This simulation exhibits the shock of the latter CME traveling within the ejecta of the first CME as observed by \citet{Lugaz2015}. The virtual spacecraft timeseries is shown in Figure~\ref{fig:waitingTime20}, with the timeseries from the individual simulations for CME1 and CME2 also shown in dashed and dotted lines, respectively. The $z$-plane density, radial velocity and southward magnetic field are shown in Figure~\ref{fig:waitingTime20hParaview}, 30 and 45\,h after launch. 

The combined simulations result in a first shock arriving at the time the CME1 shock normally would have. This is however significantly greater in magnitude and indicates that CME2 travelled faster within the reduced density wake of CME1, effectively slip-streaming before impacting and merging with CME1. The sheath densities, velocities and temperatures show a clear double-step increase and that CME2 maintains an individual front despite the impact and merge. This is also visible in Figure~\ref{fig:waitingTime20hParaview}. What is most notable is the large enhancement in the sheath density between {CME2} and {CME1}-{CME2} of +98\% from 17.8 to 34.8 m$_p$/cm$^3$. The magnetic field is significantly enhanced within this region, B$_Z$ also displaying a double peak structure, and reveals that both flux ropes now arrive several hours earlier than when individually simulated and are now located close to what would have been the high density former `sheath' region for {CME1}. This effect is visible within Figure~\ref{fig:waitingTime20hParaview} where at 30 h the B$_Z$ reveals two distinct adjacent CMEs, but at 45 h, the minima in B$_Z$ are located close to the front of the combined event. What is also noticeable is that as CME2 has travelled significantly faster, it is stretched and diffused to encompass a greater area and, also due to the greatly enhanced combined sheath in-front, shows as a reduced magnetic field intensity behind the second shock compared to when simulated individually.

The Dst timeseries for the single CME in Figure~\ref{fig:Verification_180Timeseries} demonstrate the general concept that within CMEs, both the sheath and flux rope induce Dst minima. In Figure~\ref{fig:waitingTime20}, we can now see that the CME-CME impact has resulted in two such minima combining across the same region. The minimal Dst thus reaches -267 nT, an 137 \% increase in geo-effectiveness with respect to {CME2} alone. This event would therefore be classified as a severe storm. This Dst trend is reflected in the underlying variables B$_Z$ which sees an 91\,\% increase from -11.2~nT to -21.4~nT and v$_r$ being 340~km/s greater. The enhanced density and dynamic pressure also results in the subsolar magnetopause being compressed to 5.32\,$R_E$, a decrease of 25\% from 7.11 to 5.32\,$R_E$. The region following the minima in B$_Z$ and Dst shows an extraordinary low density down to 0.15~m$_p$/cm$^3$. Here the magnetic field is stretched radially, as expected for preconditioned solar wind. 

 \subsubsection{28 hours: CME-CME Interaction}\label{sec:28h-CME-CME}
\begin{figure}
    \begin{center}
 \includegraphics[width=0.9\textwidth]{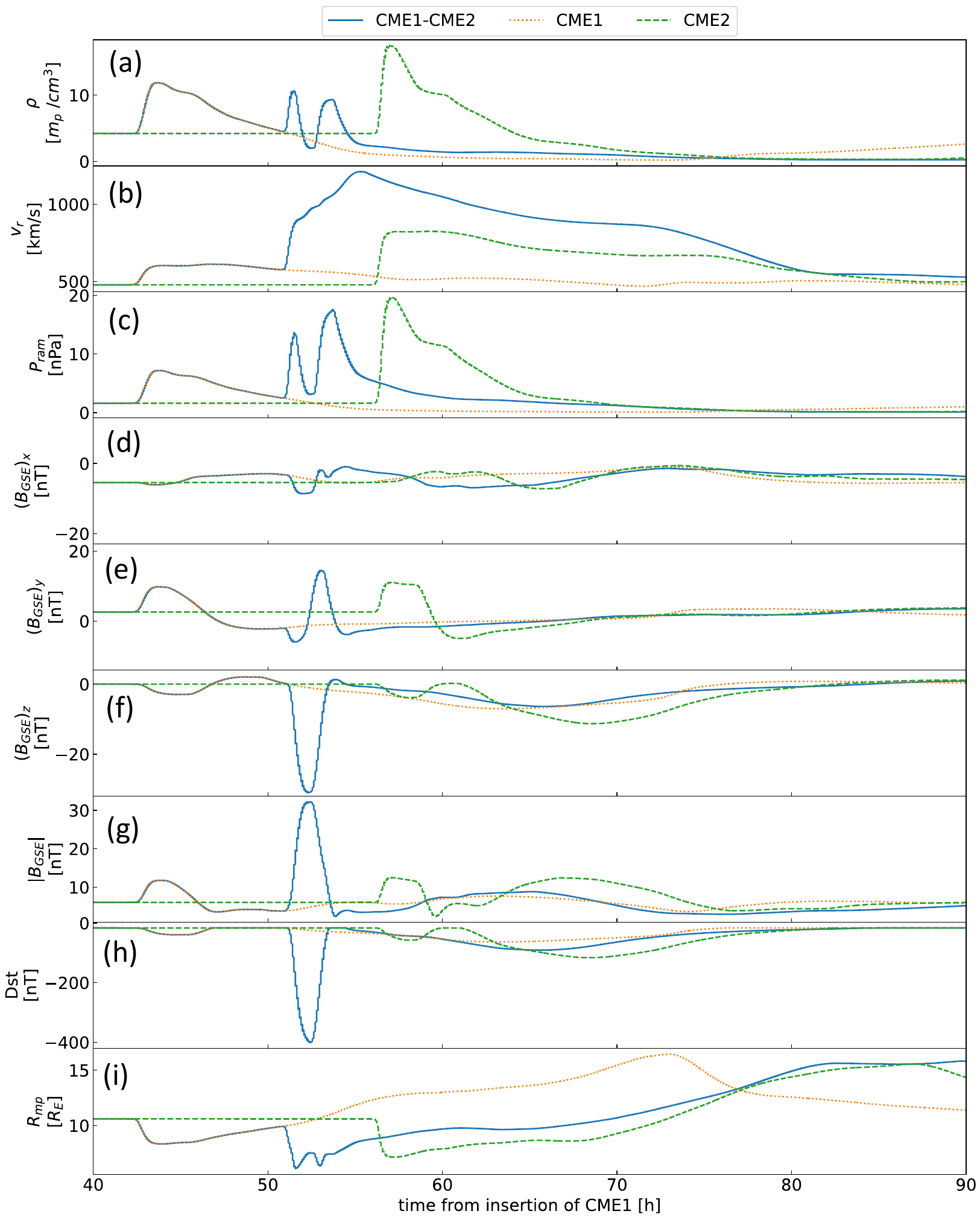} 
    \end{center}
    \caption{Virtual spacecraft data at for the CME1-CME2 interaction with waiting time of 28~h between their respective launches. This corresponded to a merge of the CME flux ropes at approximately 0.9~AU. The timeseries in solid blue, dotted yellow and dashed green refer to the signatures of CME1-CME2, CME1, and CME2, respectively.}
    \label{fig:waitingTime28}
\end{figure}

The waiting time of 28\,h results in a later merge of the magnetic flux ropes at $\sim$0.9~AU. The corresponding virtual spacecraft timeseries is shown in Figure~\ref{fig:waitingTime28} with the individual CMEs once again shown in dashed and dotted lines.

Due to the later merge, the arrival of the forward shock and sheath associated with {CME1} remains unchanged. Following this, the flux rope for {CME1} is coincident with the shock and sheath from {CME2} which itself has arrived earlier due to passing through the preconditioned solar wind behind {CME1}. This results in a double peak in the density and a highly compressed magnetic field signature. The most striking feature of this virtual spacecraft timeseries is thus the strong and sharp increase in B$_Z$ due to strong decrease in the $(B_{GSE})_z$. The minimal $(B_{GSE_z})$ for the {CME1}-{CME2} timeseries is -31~nT representing a 174 \% change with respect to the {CME2} timeseries. The minimal Dst for the {CME2}-{CME1} timeseries is predicted to be -401~nT, again representing a 245\% change and an even more severe storm than in the 20~h waiting time scenario examine previously. The subsolar magnetopause position is compressed by 14\% from 7.11\,$R_E$ for {CME2} to nearly 6.09\,$R_E$ for the combined {CME1}-{CME2}. The subsolar magnetopause is not as compressed as in the {20~h} merge due to a low peak in dynamic pressure. This highlights the need for differing measures of geo-effectiveness which represent difference facets of geomagnetic storms at the Earth.

\subsubsection{Overarching Trends}
\begin{figure}
    \begin{center}
            \includegraphics[width=0.5\textwidth]{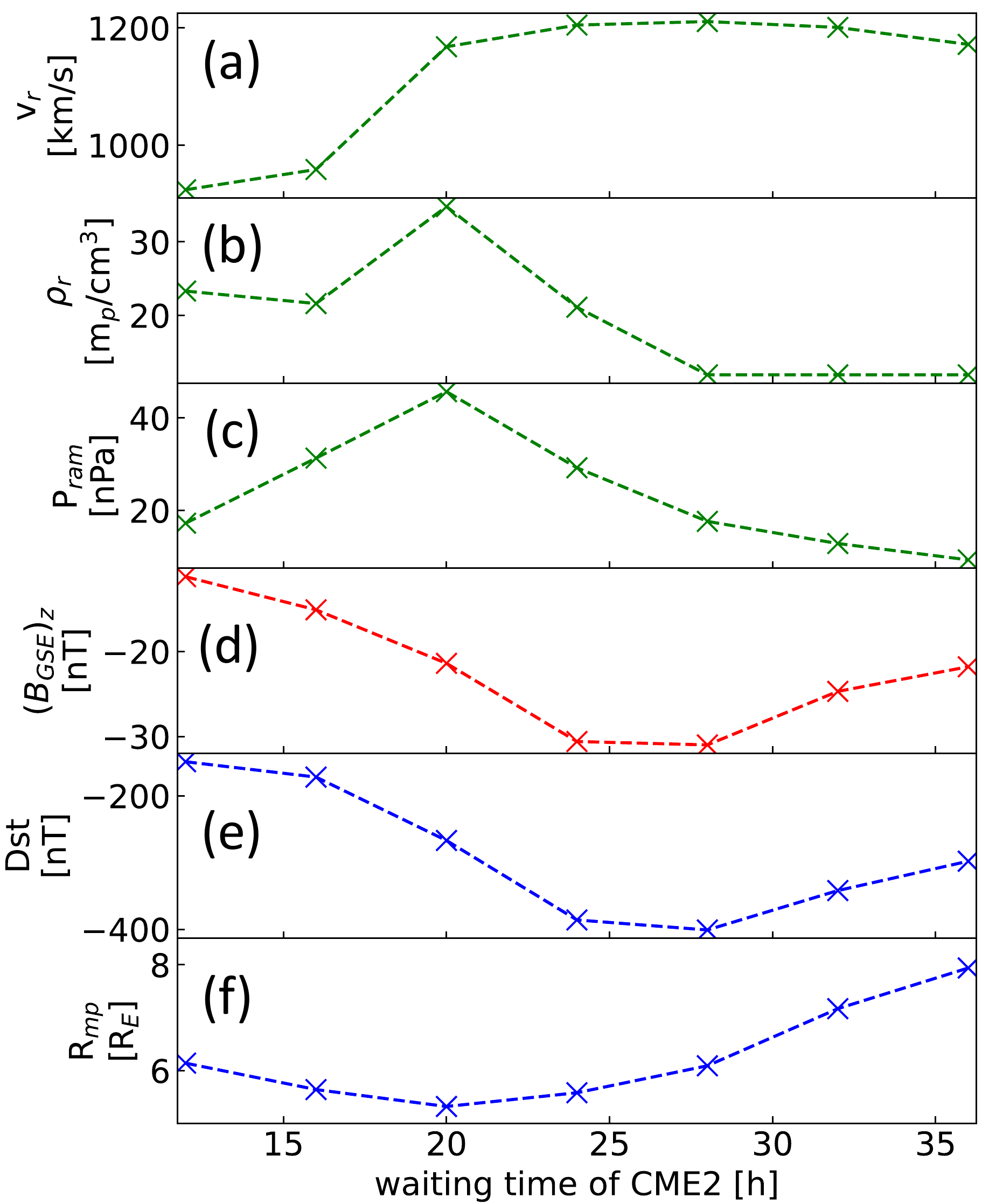}
    \end{center}
\caption{ Maximal values for the range of waiting times CME1 and CME2 are simulated for. (a) shows v$_r$, (b) shows the density, $\rho_r$, (c) the dynamic pressure, P$_{ram}$, (d) $(B_{GSE})_z$, (e) the predicted Dst and (f) the predicted subsolar magnetopause standoff distance, R$_{mp}$.}\label{fig:MaxWaitingTime}
\end{figure}

To draw conclusions about the relationship between geo-effectiveness and waiting time, the absolute maxima within the {CME1}-{CME2} simulations for selected physical variables at 1~AU are summarised in Figure~\ref{fig:MaxWaitingTime}.\\
To identify the type of event we distinguish between the interaction of the CME sheath regions and the slower traveling CME flux rope centres. Here the clearly identifiable merge of sheath regions is taken as the defining characteristic of an interaction. The range of waiting times of 12~h, 16~h, 20~h and 24~h correspond to full merges of the CME sheath regions at $\approx$0.5~AU, 0.7~AU, 0.9~AU and just after Earth, respectively. All other runs do not exhibit a merger within the simulation domain, extending up to 1.1~AU. Thus one may identify the first three simulations, 12-20~h, as collision events and the last two, 32-36~h, as mostly preconditioning cases. The events in the middle of these exhibit features of both regimes.

The maximal and delta velocity show a plateau of a maximal velocity of approximately 1200~km/s and a relative increase from the faster CME2 of 400~km/s, due to the low densities behind CME1 and thus lower drag forces acting on CME2. Short waiting times do not appear to allow sufficient time for propagation of {CME1} to impact on the speed of propagation of {CME2}. The 36~h waiting time case already shows a recovery of the solar wind speed toward its nominal state.

The maximal density shows an increase up to the 20~h waiting time case due to the sheaths merging and becoming the most concentrated just ahead of the Earth, as is discussed in Section~\ref{sec:28h-CME-CME}. This translates into the most compressed magnetopause position. After 20 hours the maximal density rapidly falls away due to the CME sheath regions no longer fully merging. The minimum southward magnetic field component shows a smooth depression centred at 28~h. This is due to the 28~h case, as discussed in Section~\ref{sec:28h-CME-CME}, featuring a merge of the magnetic flux ropes close to 1~AU which produces a highly concentrated B$_Z$. This effect must also occur during earlier mergers but the lack of such a strong B$_Z$, and indeed two distinct CME signatures in Figure \ref{fig:waitingTime20} can be attributed to the conversion of magnetic energy into kinetic and thermal components \citep{Scolini2020} or the relaxation of this concentrated flux due to the elastic nature of CME-CME interactions previously identified by \cite{Shen13}. For longer waiting times of 26 hours and beyond, a recovery to the solar wind normal starts to appear, see also in~\citet[][Figure 5]{Desai2020} for the extension to this trend where the initial CME has no impact on the subsequent event for a sufficiently long waiting time between their eruptions.

The maximum B$_Z$ translates into a maximum Dst close to the 28~h case, with a Dst of -401~nT, which would be a great or extreme storm. It must be noted here that the Dst relationship used is only linear in velocity and is not a good predictor for high velocity events as investigated here. The reflected trends in the underlying variables, however, appear to justify this characterisation of an extreme event.

\subsection{CME-CME Stage 3: Handedness}\label{sec:Handedness}
The third and final stage of this parametric study concerned itself with the handedness, $H$, the results of which are displayed in Table 2. The handedness defines the sign of the $B_{\phi}'$ component. A positive handedness $H = +1$ results in the magnetic field lines turning counterclockwise looking down on the spheromak and a negative handedness correspondingly a clockwise rotation. The intention of this stage was to test whether a particular handedness combination would result in altered structural properties of the CMEs at 1~AU and therefore potentially a more geo-effective interaction. It must be emphasized that the initial $(B_{GSE})_z$ component of the spheromak is kept constant. A total of four simulation runs were run for the 20 hour case to capture all $H_1 = \pm 1$ and $H_2 = \pm 1$ combinations, referring to {CME1} and {CME2} respectively, and each CME is also simulated individually. This 20~h case captured a balance of an extended interaction and significantly enhanced B$_Z$ and geo-effectiveness. In the following we shall refer to each run as H:[$H_1$, $H_2$] = H:[$\pm$1, $\pm$1].

The difference between the runs is most apparent in tabular form of the maximal values, {as we show in} Table~\ref{tab:HandednessSummary}. Even though all runs are initialized with the same (B$_{GSE}$)$_z$ structure, a large variation in the signatures at Earth is found. A clear trend from positive to negative handedness is identified and the handedness of the first CME has a higher influence than the handedness of the second CME. The maximal southward magnetic field varies between -12.7 to -21.8~nT. This is a 71\% difference over which the dynamic variables show no significant trend. The resulting Dst consequently also exhibits this same trend, increasing from -163 to -275~nT, a 68\% change. Note again that the minimum in Dst are found in sheath/shock regions due to the merge of the high density sheath and subsequent flux ropes. These differences in geo-effectiveness would already change the classification of the geomagnetic storm from strong to severe. It is thus a significant finding that the handedness of spheromak in CME-CME interactions can have such a dramatic effect on the geo-effectiveness.

\begin{figure}[ht]
    \begin{center}
            \includegraphics[width=0.6\textwidth]{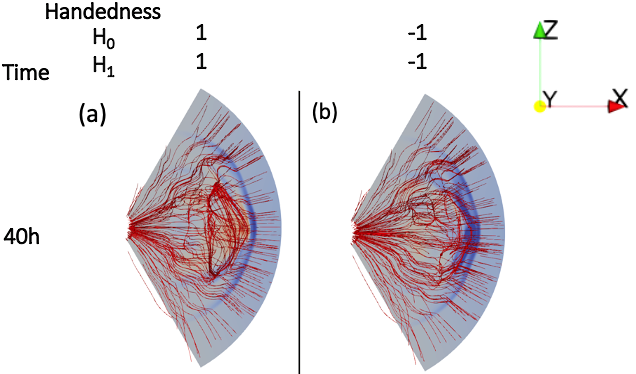}
    \end{center}
    \caption{Meridional plane profiles and selected magnetic field lines for the collision of CME1 and CME2, 40~h after CME1 initialisation. Each handedness variation is shown (a-b) for a waiting time of 20~h and tilt angles of $\tau$ = 180$^{\circ}$. 
    }    \label{fig:h_yprofilesFlowlines}
\end{figure}

To investigate the cause of this dramatic impact, Figure~\ref{fig:h_yprofilesFlowlines} shows the $y$-plane profiles of $B_{\phi}$ and the 3D magnetic field lines projected onto that plane. It only includes the most and least geo-effective cases, H:[+1,+1] and H:[-1,-1] respectively at 40~h. This figure indicates the potential cause for this significant difference in geo-effectiveness. The H:[+1,+1] case exhibits a greater winding of magnetic field lines in the CME-CME region in comparison to the  H:[-1,-1] where in this case the CME structure appears to merge with the joint sheath region. We therefore propose that this behaviour may be explained as follows. For a spheromak of $\tau = 180^{\circ}$, a handedness of $H=+1$ means that the internal magnetic field lines in the toroidal direction flow in the same sense as the Parker spiral on the radially outward propagating side. This is also referred to as an inverse CME as defined by \citet{Low2002}. This results in a greater conservation of field lines within the spheromak. This preferentially directed toroidal flux lends stability to the magnetic ejecta and results in a prolonged compression of field lines. A negative handedness, $H=-1$, on the other hand leads to field lines of spheromak and Parker spiral flowing in the opposing directions on the outward radial side. This results in greater erosion of the spheromak due to magnetic reconnection \citep{McComos94,Schmidt2003,Gosling2005,Gosling2007} and explains the de-facto erosion of magnetic flux for the H:[-1,-1] case.
The above explanation is supported by the observation that the first CME's handedness has a larger effect than that of the second one, see the progressive trend in Table~\ref{tab:HandednessSummary}. This is because only the first CME strongly interacts with the Parker spiral as it is compressed by the subsequent ejecta, whereas the second CME travels within the preconditioned solar wind where the magnetic field is more stretched in the radial direction. This is consistent with the single CME simulations where only the faster CME has an altered B$_Z$ at 1~AU and interestingly also an altered velocity. The results of the single CME simulations are therefore also consistent with high-resolution 2.5D MHD simulations \citep{Hosteaux2019,Hosteaux21} which found that the CME polarity had an effect on the CME flux and that increased velocities resulted in greater erosion.

\section{{Discussion} \& Conclusions}\label{sec:Conclusion}
This study has conducted a parametric evaluation of successive interacting coronal mass ejections in a representative heliospheric environment. We employed a 3D heliospheric MHD model~\citep{Desai2020} and implemented a force-free spheromak CME description~\citep{Verbeke2019}. The force-free implementation was first verified within a static simulation representative of 0.1~AU and then the CME was propagated to 1~AU where a virtual spacecraft demonstrated the modelled CME fulfilled the criteria for a magnetic cloud~\citep{Burlaga1981}.

We then conducted a parametric evaluation of successive interacting coronal mass ejections using CME parameters defined by~\citet{Scolini2019} for the two eruptions identified on 13-14 July 2012. Two CMEs were thus simulated, a slow and then fast one, and the collision and the resulting characteristics at 1~AU evaluated using timeseries from a virtual spacecraft. As over 90 \% of CMEs have only been observed by a single spacecraft, this provided a clear interpretation of the simulation results in context with the literature, whilst the global simulations provided context to explain these timeseries. The results at 1~AU were thus used to assess the geo-effectiveness through quantification of the relevant solar wind variables and an empirical measure of the Disturbance Storm Time index and subsolar magnetopause standoff distance.

The CME tilt angle was first varied and the B$_Z$ component specified during initialisation was found to be well conserved within this idealised environment with a tilt angle of 180$^\circ$, thus providing the most geo-effective event at 1~AU. The launch time between the successive CMEs was then varied to produce full CME-CME interactions starting at 0.5~AU out to beyond the Earth, the latter of which represents a case of pure solar wind preconditioning. The various merges produced a diverse range of geophysical impacts with different solar wind variables peaking for merges at different launch times. For example, a waiting time of 20~h produced a collision at 0.5~AU and the prolonged interaction resulted in the maximum dynamic pressures due to the merge of the CME sheath regions. This resulted in the minimum estimated subsolar magnetopause position of just 5.12~R$_E$. A waiting time of 28~h produced a less complete merge close to the Earth. This however resulted in the two spheromaks colliding near to the Earth which produced the most concentrated and greatest B$_Z$ signature which translated into the most geo-effective event according to the Dst estimates. The Dst here was predicted to be less than -400\,nT and, due to similarly amplified underlying solar wind variables, led us to classify this as an extreme event. These results thus demonstrate how two CMEs, with an optimal waiting time between them, can convert CMEs which would typically result in only moderate storms, into severe and extreme storms. The final stage then took the 20~h case, due to its prolonged interaction, and varied the handedness, also referred to as chirality, of the CMEs to determine whether this induced any effect at 1~AU. While the handedness only had a negligible effect for a single CME, the CME-CME interaction simulations showed a clear trend of increasing B$_Z$, and thus Dst, for handedness changes from H=+1 to H=-1. This was attributed to a positive handedness resulting in the spheromak field lines pointing in the opposite direction to the Parker spiral which resulted in erosion of the CME's magnetic structure {due to magnetic reconnection.

Magnetic reconnection in the solar wind was first reported to exist upstream of CMEs \citep{McComos94,Gosling1995,Schmidt2003}, and indeed potentially at the interface between two magnetic clouds and further thin current sheets associated with large changes in the magnetic field direction \citep{Gosling2005}. Since then, reconnection associated with CMEs has been identified by several studies \citep[e.g.][]{Ruffenach12,Lavraud14}. The findings of this study build upon these results and highlight that within CME-CME interactions, this process can potentially have an enhanced effect on the global CME structures and thus CME geo-effectiveness at the Earth. The explanation that we propose here highlights that early identification of the chirality of a given CME in the solar atmosphere~\citep[e.g.][]{Deforest17,Palmerio2017} presents a potentially significant diagnostic for identifying geo-effective events. 
It is important to note, however, that magnetic reconnection in ideal MHD codes is dependent on numerical resistivity. While this has been shown to be appropriate for simulations of the global dynamics of the Earth's magnetosphere \citep[][]{Toffoletto09} which exists within a comparable range of plasma betas and Lundquist numbers to the CME simulations conducted herein, such an approximation of kinetic physics might well influence the results. Further studies of the microphysics of the interaction between magnetised solar wind structures and the ambient IMF are therefore required to further understand the large-scale trends discovered herein.}

This effect {of large-scale magnetic field erosion} appeared dominant for the leading CME which predominantly interacted with the Parker spiral. The final stage of this study thus found that the handedness of a CME within a CME-CME event can alter the geo-effectiveness from a moderate to severe event. 

This parametric study of CME-CME interactions demonstrates how two moderate CME events, which individually induce an estimated Dst of just -63~nT can combine to increase their characteristics and geo-effectiveness. The variety of possible merges examined induced a range of Dst estimated up extreme cases of less than -400 nT. The complexities of the interactions between the Sun and the Earth, highlight the difficulties in providing accurate long-term space weather forecasts based on the solar corona and highlight the need for self-consistent physics-based modelling approaches to capture the magnetised interactions within our heliosphere. 

\newpage
\section*{Acknowledgements}
GJK carried out this masters research project at Imperial College London. RTD acknowledges an STFC Ernest Rutherford Fellowship ST/W004801/1, and NERC grants NE/P017347/1 and NE/V003062/1. EED is supported by NASA grant 80NSSC19K0914. JPE was supported by NERC grant NE/V003070/1. SP acknowledges support from the projects C14/19/089  (C1 project Internal Funds KU Leuven), G.0D07.19N  (FWO-Vlaanderen), SIDC Data Exploitation (ESA Prodex-12), and Belspo project B2/191/P1/SWiM. This study used the Imperial College High Performance Computing Service (doi: 10.14469/hpc/2232).

%% pdfsP sample631.tex

\bibliography{sample631}{}
\bibliographystyle{aasjournal}



\newpage

\begin{table}[ht]

 \caption{Parameters chosen as the base CME for LFFS model in this paramedic study. Parameters in light grey cells are fixed throughout the study. Parameters in bold have been adapted directly from observational estimations by~\cite{Scolini2019} for realistic modelling. All plain cells are varied throughout \emph{Stages 1-3}. $\dagger$ indicates a derived quantity not directly inputted, i.e. implied by other parameters. }
    \label{tab:ParameterStudy}

    \centering
        \footnotesize
         \begin{tabular}{lllc} 
         
         \hline\hline
                                            & Variable                                           & CME1               &  CME2 \\ 
           \hline
                                            & \scriptsize{Insertion time of CME [in h]} & 0                 &   12-36 \\
    \rowcolor{LightGrey}
                                            & \scriptsize{\textbf{Initial radius} ${r_0}$ [in $R_S$]}       & \textbf{10.5}               & \textbf{10.5}  \\
    \rowcolor{LightGrey}
            \scriptsize{Initial position}   & \scriptsize{Radial $r_{CME}$}                      & 0.1 AU + $r_0$     &   0.1 AU + $r_0$\\
    \rowcolor{LightGrey}
                                            & \scriptsize{Polar $\theta_{CME}$}                  & 90$^\circ$                &  90$^{\circ}$ \\
    \rowcolor{LightGrey}
                                            & \scriptsize{Azimuthal $\phi_{CME}$}                & 0$^{\circ}$                 &  0$^{\circ}$    \\
    \rowcolor{LightGrey}
            \scriptsize{Velocity}           & \scriptsize{Total $v_{3D}^\dagger$ [in km/s]}              & 723                &   1723  \\
    \rowcolor{LightGrey}
                                            & \scriptsize{Radial $v_{r}$ [in km/s]}              & 500                & 1500   \\
    \rowcolor{LightGrey}
                                            & \scriptsize{\textbf{Magnetic expansion} $ {v_{exp}^\dagger}$ [in km/s]}& \textbf{223}                &  \textbf{223}  \\
    \rowcolor{LightGrey}
           \scriptsize{Magnetic field}            & \scriptsize{\textbf{Field strength} $ {B_0}$ [in nT]}          & \textbf{1400}               &  \textbf{1400}  \\
                                            & \scriptsize{Tilt angle $\tau_{CME}$}               & 0-270$^{\circ}$             &  0-270$^{\circ}$   \\
                                            & \scriptsize{Handedness $H$}                        & +1 (or -1)          &  +1 (or -1)  \\
          \hline
          
       \end{tabular}
       \label{Table1} 
\end{table}

\begin{table}[ht]
    \centering
    \caption{Summary of extrema of handedness CME-CME runs, accompanied by single CME runs of CME1 and CME2 for comparison.}\label{tab:HandednessSummary}
    
        \footnotesize
         \begin{tabular}{cc|ccccc} 
         
         \hline\hline
                \multicolumn{2}{c}{Handedness}   & \multicolumn{5}{c}{Max/Min values}                                       \\ 
           \hline
                $H_1$       &   $H_2$            & $(\rho)_{max}$  [m$_p$/cm$^3$]   & $(v_r)_{max}$ [km/s]  & $((B_{GSE})_z)_{min}$ [nT]    & $(Dst)_{min}$ [nT] & $(R_{mp})_{min}$ [R$_{mp}$] \\
           \hline
                1           &   1                & 33.7                       & 1144           &   -21.8                 &  -275      & 5.23\\
                1           &  -1                & 33.0                       & 1166           &   -20.1                 &  -243      & 5.38\\
               -1           &   1                & 34.4                       & 1132           &   -17.6                 &  -212      & 5.59\\
               -1           &  -1                & 32.4                       & 1180           &   -12.7                 &  -163      & 5.80\\
          \hline
                1           &   -                & 12.0                       & 609.9          &   -6.93                 &  -62.5     & 8.34\\
                -1          &   -                & 12.8                       & 611.8          &   -6.97                 &  -62.0     & 8.24\\
                -           &   1                & 17.3                       & 866.9          &  -10.7                  &  -111.0    & 7.06\\
                -           &   -1               & 17.7                       & 820.2          &  -12.0                  &  -119.0    & 7.15\\    
          \hline
       \end{tabular}
\end{table}

\end{document}